\documentclass[sigconf,authorversion,nonacm]{acmart}

\AtBeginDocument{%
  \providecommand\BibTeX{{%
    \normalfont B\kern-0.5em{\scshape i\kern-0.25em b}\kern-0.8em\TeX}}}

\usepackage{multirow}
\usepackage{hyperref}

\setcopyright{none}
\begin{document}
\fancyhead{}

\title{RGRecSys: A Toolkit for Robustness Evaluation \\ of  Recommender Systems}

\author{Zohreh Ovaisi}
\email{zovais2@uic.edu}
\affiliation{%
\institution{University of Illinois at Chicago}
}

\author{Shelby Heinecke}
\email{shelby.heinecke@salesforce.com}
\affiliation{%
\institution{Salesforce Research}
}

\author{Jia Li}
\email{jia.li@salesforce.com}
\affiliation{%
\institution{Salesforce Research}
}

\author{Yongfeng Zhang}
\email{yongfeng.zhang@rutgers.edu}
\affiliation{%
\institution{Rutgers University}
}

\author{Elena Zheleva}
\email{ezheleva@uic.edu}
\affiliation{%
\institution{University of Illinois at Chicago}
}

\author{Caiming Xiong}
\email{cxiong@salesforce.com}
\affiliation{%
\institution{Salesforce Research}
}

\begin{abstract}
Robust machine learning is an increasingly important topic that focuses on developing models resilient to various forms of imperfect data. Due to the pervasiveness of recommender systems in online technologies, researchers have carried out
several robustness studies focusing on data sparsity and profile injection attacks. Instead, we propose a more holistic view of \textit{robustness} for recommender systems that encompasses multiple dimensions - robustness with respect to sub-populations, transformations, distributional disparity,  attack, and data sparsity. While there are several libraries that allow users to compare different recommender system models, there is no software library for comprehensive robustness evaluation of recommender system models under different scenarios. As our main contribution, we present a \textit{robustness} evaluation toolkit, Robustness Gym for RecSys (RGRecSys),\footnote{https://www.github.com/salesforce/RGRecSys} that allows us to quickly and uniformly evaluate the robustness of recommender system models. 
\end{abstract}



\maketitle
\section{Introduction}

Recommender systems are a core component of many online personalization systems that assist users to discover their favorite items among large choice sets. They have garnered considerable interest among researchers due to their inevitably high impact on both users' experience when searching for their item of interest, as well as content providers who seek enough exposure and visibility for their products. Traditionally, recommender systems were built and studied under simple yet unrealistic assumptions, such as \emph{i.i.d.} assumptions (training and testing data being independent and identically distributed) as well as noiseless and abundant data. Recent studies relax these assumptions and focus on developing models under more challenging, yet realistic scenarios where data fed into recommender systems are maliciously attacked \cite{zhang2020gcn}, sparse \cite{zhou2020interactive}, and bias \cite{joachims2017unbiased,ovaisi2021propensity, ovaisi2020correcting}. However, there are other factors to consider when evaluating robustness. For instance, some of the user and item features may be corrupted (transformation), or the \emph{i.i.d.} assumption of training and testing data may be violated (distributional shift) \cite{cao2016non,li2021user}. The performance of recommender systems over-relying on unrealistic assumptions can be highly degraded. 

Here, we propose a holistic definition of robustness for recommendation systems that encompass and formalize several perspectives of robustness - robustness with respect to sub-population, transformations, distributional disparity, attack, and sparsity.
Moreover, with the fast growth of recommender system models, it is important to develop libraries to allow researchers to easily reproduce and evaluate different models and test their robustness in aforementioned directions. Thus, in addition to our conceptual contribution, we present a robustness evaluation toolkit for recommendation systems, Robustness Gym for RecSys (RGRecSys), which allows us to quickly and uniformly conduct a comprehensive robustness evaluation for recommendation system models. In what follows, we first discuss related work before proceeding to describe the library capabilities in detail. We then present a case study that demonstrates the features of our library. We conclude with our final reflections and future directions. 

\section{Related Work} 
\label{sec:related}
With the fast growth of recommender system models, it becomes essential to develop a library with built-in state-of-the-art models to allow researchers for a fair and comprehensive comparison of such models. To this end, a large number of recommender system libraries have been developed recently, including Mahout\footnote{https://mahout.apache.org}, Duine\footnote{http://www.duineframework.org}, Cofi\footnote{http://www.nongnu.org/cofi/}, LensKit \cite{ekstrand2011rethinking}, MyMediaLite\footnote{http://www.mymedialite.net}, PREA \cite{lee2012prea}, LibRec \cite{guo2015librec}, and RecBole \cite{zhao2020recbole}. Mahout, Duine, and Cofi are libraries that develop only memory-based models
\cite{lee2012prea}. LensKit proposes a library for collaborative filtering models in Java.  MyMediaLite and PREA are more advanced packages. MyMediaLite is a  library based on C\# that addresses rating prediction and item prediction from implicit feedback for collaborative filtering models. PREA is a library providing an evaluation for memory-based and matrix-factorization recommender systems in Java. However, these libraries are barely active for further improvement. LibRec introduces a Java-based library that covers more built-in recommender system models such as general, context-aware, social, and graphical models. Finally, RecBole is a unified PyTorch-based framework that reproduces 73 models in several categories of general recommendation, sequential recommendation, context-aware recommendation, and knowledge-based recommendation on 28 benchmark datasets. All of these libraries only evaluate models under strong assumptions like \emph{i.i.d.} assumptions as well as perfect and abundant data. However, data fed into the recommender systems are often sparse (especially if they are explicit feedback from users), and also inaccurate. Thus, it is vital to design a library that evaluates models under such more challenging scenarios. Our library is highly inspired by the work of \cite{goel-etal-2021-robustness} that evaluate Natural Language Processing (NLP) models under sub-population, transformation and attack. RGRecSys evaluates recommender system models' robustness against data sub-population, transformation, distributional shift, attack and sparsity.

\section{Our Framework}
\label{Framework}

\subsection{Library Architecture and Contents}

As RecBole \cite{zhao2020recbole} is the state-of-the-art recommender system library with a broad set of models and advanced features, we use RecBole built-in models to demonstrate RGRecSys usefulness. RecBole adopts Pytorch in the entire library and proposes a unified framework with data, model, and evaluation modules. This library has a comprehensive set of models on different categories of general, sequential, context-aware, and knowledge-based recommendation. Its general and extensible data structure ease the process of adding new models to this library and gives enough flexibility to its users to  set up experimental environments such as hyper-parameters and splitting criteria. RGRecSys has a robustness evaluation module that allows us to conduct unified and comprehensive robustness evaluations on recommender system models. Below, we discuss RGRecSys robustness features in detail.

\subsection{Subpopulation}
Most existing recommender system libraries report the performance metrics averaged on overall users and items. However, a single high-performance metric does not guarantee that the model has the same high performance for a sub-group of users or items. For instance, a recommender system might have a great performance averaged over all users but have a poor performance on a sub-group of users like females or people of a certain race. With the recent vital importance of fair recommender systems \cite{singh2018fairness}, it is critical to report the performance for a slice of users or items. Our library gives this flexibility to evaluate the model performance for any sub-group of interest, such as users of a particular feature, users' activeness based on their numbers of interactions as well as users' critiques based on their rating score. That is, given a trained model, the library can perform slicing on test data to conduct fine-grained evaluation on models and evaluate their robustness to slicing.
\subsection{Distributional shift}
Many recommender system models are designed based on the assumption that training and testing data have independent and identical distribution. However, this \emph{i.i.d.} assumption is often violated in real scenarios, which leaves the great performance of many existing models questionable \cite{shen2021towards}. This is especially true for recommender systems as the training data is collected using the existing production system, the distribution of the data will be different when the newly developed model is deployed online \cite{cao2016non}. RGRecSys allows us to evaluate recommender systems models under distributional shift. To this end, RGRecSys first provides the users of our library with the distribution of training data based on user features, and then allow them to manipulate the testing data distribution by sampling it so that its distribution becomes different from training data. For instance, our library users can decide the female and male ratio in testing data that is highly different from training data.
\subsection{Transformations}

Most recommender system models need to access user and item features to provide users with a set of recommended items. Such data can be either collected by asking the users and content providers to complete a profile that includes some information about users or items, or the recommender system may automatically extract such data, e.g., from user's reviews. However, this data might be corrupted either due to misleading information, or any error that can arise when recommender models aim to extract them, or by some form of malicious attack. For instance, a user might misinform the system about his location by connecting through VPN, a person who aims to sell an item on the eBay platform might not mention certain classes (categories) that the item can fall into, a recommender system using machine learning algorithm may misjudge when trying to extract some information about users and items in a review comment or photo, or malicious attacks can corrupt such features. Thus, it is more realistic to assume that user or item features are inaccurate to some extent. RGRecSys allows us to evaluate models under transformation on user or item features by giving them the flexibility to choose feature(s) they want to transform, and the severity of this transformation. This transformation can be either random where the feature value can take any random value, or structured where the feature value is within a certain distance of its true value. The first one is indicative of a scenario where the wrong user information or an attack misleads the system and the second one is indicative of models' inaccuracy when extracting features (e.g., user age). Note that transformation on features only occurs on testing data so as to evaluate if a model is robust to such changes. This form of transformation is not applicable to general recommender systems that do not use features in prediction process. As some users in training data may appear in testing data too, this transformation translates to the case where some users have corrupted feature value in testing data while their feature in training data is remained uncorrupted. This is true because model training is time-consuming, and real-world recommender system models are not re-trained very frequently. Thus, the time period between re-training model is long enough for some users' and items' features to be attacked or unintentionally changed due to data processing errors. We expect models that overly rely on certain users and items feature(s) would suffer when such features are corrupted.
\subsection{Attack}

With the huge economic impact of recommender systems, they are highly prone to be attacked with the purpose of increasing or decreasing some items rankings. Thus, it is critical to evaluate recommender system models performance when dealing with malicious attacks. While there are many types of attack designed to degrade recommender system performance, RGRecSys allows us to evaluate the models under Cross-Site Request Forgery (CSRF) attack \cite{barth2008robust}, where the attacker causes the victim user to carry out an action unintentionally. For instance, the user may unwillingly change his ratings, leading to corrupted interaction data in training dataset. RGRecSys allows us to decide the severity of attack by deciding what fraction of interaction would be corrupted.

\subsection{Sparsity}
Data fed into recommender systems usually contain explicit or implicit feedback from users, e.g, ratings or clicks. Such data are usually highly sparse, and recommender systems performance is known to be degraded when fed by such sparse data. Our library allows its users to compare the robustness of different models under sparse data by randomly removing a fraction of user interaction data. The sparsity level as well as which users (based on their activity) one may want to drop interactions of, can be freely chosen.

\section{Case Study}
Here, we demonstrate experimental results on robustness evaluation on a variety of general and context-aware recommendation models. The general recommendation models featured in the examples below are Pop (popular items), NeuMF,  LightGCN, ItemKNN, and SpectralCF \cite{he2017ncf, he2020lightgcn, aiolli2013efficient, zheng2018spectral}, whereas we chose DCN, DSSM, AFM, LR, and FM \cite{wang2017deep, huang2013learning, xiao2017attentional, richardson2007predicting, rendle2010factorization} as the context-aware recommendation models for demonstration. We show example results on one data set, MovieLens 100k \cite{harper2015movielens}. This dataset contains $100,000$ ratings $\in [1,5]$ from $943$ users on $1,682$ movies. For equal comparison, the train, validation, and test split are consistent throughout all examples (via the same random seed), and are ratio-based with $80\%$, $10\%$ and $10\%$ of data to be train, valid and test respectively. We use the default model hyperparameters as defined in \cite{zhao2020recbole}. To define robustness tests, one must simply create a \texttt{robustness config} dictionary using the appropriate keys and values \footnote{See our documentation at https://www.github.com/salesforce/RGRecSys for more details.}. 
We emphasize that due to the limited trials and dataset used in the examples below, these examples alone are not sufficient to draw strong conclusions. Instead, the purpose is to highlight the capabilities of our library. We leave extensive experiments and insights about the robustness of various types of recommendation models to future work. 
\subsection{Subpopulation with Respect to User Gender}
\begin{figure}[h]
    \centering
    \includegraphics[width=.49\linewidth]{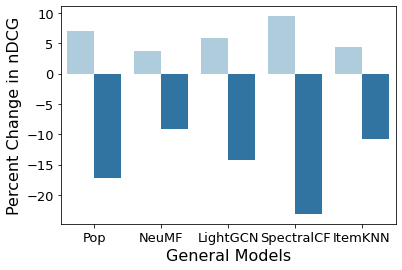}
    \hfill
    \includegraphics[width=.49\linewidth]{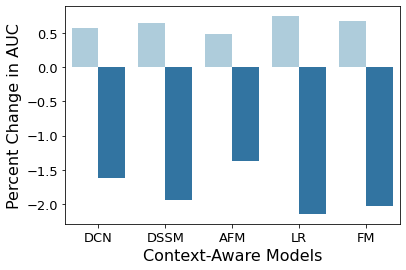}
    \caption{The percent change in nDCG and AUC between the original test set and the subpopulation of test set consisting of female (dark blue) and male (light blue).}
    \label{fig:subpopulation_women}
\end{figure}

For many applications, model performance on subpopulations, such as those defined by gender, are important to consider. In this example, we create a subpopulation of the test set consisting of users who identify as females (dark blue) and males (light blue). We first train our general and context-aware recommendation models on the same training data and evaluate each on both the original test set containing all users and the subpopulation. Note that the general recommendation models do not leverage user features, in contrast to context-aware models, but we still use these features for filtering the test set. For brevity, we show the percent change in nDCG and AUC between the original test set and the subpopulation for general and context-aware models in Figure \ref{fig:subpopulation_women} respectively. While more insight in data statistics is required for any reasoning, it is shown that the system performs much worse for females across models. Also, SpectralCF and LR exhibit the largest performance gap for females among the general and context-aware models respectively.

\subsection{Distributional Shift with Respect to User Gender}
\begin{figure}[h]
    \centering
    \includegraphics[width=.49\linewidth]{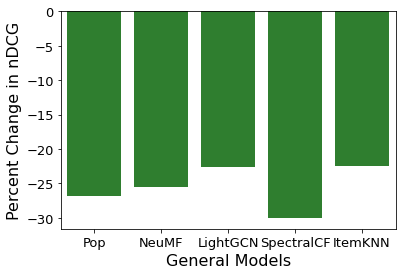}
    \hfill
    \includegraphics[width=.49\linewidth]{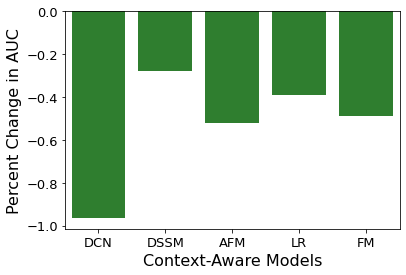}
    \caption{The percent change in nDCG and AUC between the original and distributionally shifted test set.}
    \label{fig:shift_women}
\end{figure}

To have a different distribution between training and testing data, we sample the testing data based on user gender. The training distribution contains $74\%$ male and $26\%$ female. We set the test set distribution to $50\%$ male and $50\%$ female. Figure \ref{fig:shift_women} shows that while all models performance degraded, ItemKNN and DSSM are the least vulnerable among general and context-aware models, respectively.

\subsection{Transformation with Respect to User Age}
\begin{figure}[h]
    \centering
    \includegraphics[width=.6\linewidth]{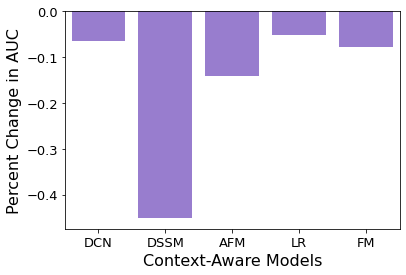}
    \caption{The percent change in AUC between the the original and transformed test set.}
    \label{fig:transformation}
\end{figure}

Here, we apply our structured transformation tool to perturb each user's age in the test set by no more than 10\% deviation from its true value and evaluate a variety of context-aware models. As shown in Figure \ref{fig:transformation}, DSSM exhibits the most vulnerability to this perturbation.

\subsection{Attack on Ratings History}
\begin{figure}[h]
    \centering
    \includegraphics[width=.6\linewidth]{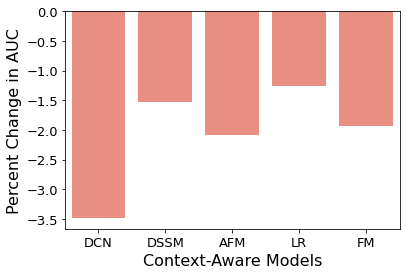}
    \caption{The percent change in AUC between models trained on the original and attacked train set.}
    \label{fig:attack}
\end{figure}

In this example, we implement an attack to the rating history using our library API for context-aware models. We randomly select 10\% of interactions from the training data set and change the rating to a random value to mitigate the effect of CSRF attack \cite{barth2008robust}. We train all models on both the original training set and the attacked training set and evaluate on the same test sets.
As depicted in Figure \ref{fig:attack}, the performance of all context-aware models are shown to be degraded, where LR is the least affected model.

\subsection{Sparsifying Rating History}
\begin{figure}[h]
    \centering
    \includegraphics[width=.49\linewidth]{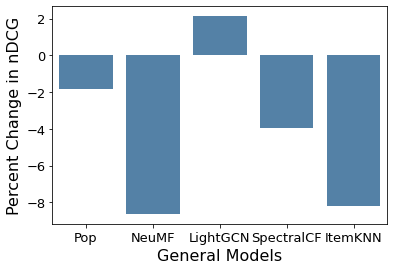} 
    \hfill
    \includegraphics[width=.49\linewidth]{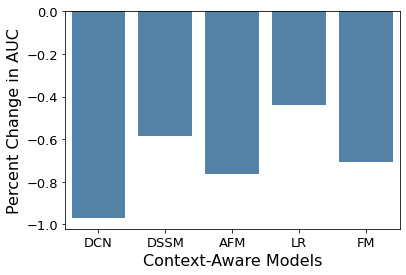}
    \caption{The percent change in nDCG and AUC between the model trained on the original and the sparsified train set.}
    \label{fig:sparsity}
    
\end{figure}

Here, we consider the sensitivity to sparse data across several general and context-aware recommendation models. For each user, we uniformly and randomly remove 25\% of their interactions. We train all models on both the original training data and the sparsified training data and evaluate on the same test sets. As shown in in Figure \ref{fig:sparsity}, NeuMF and DCN are most vulnerable models to sparsity. 

\section{Conclusion and Future Work}
In this work, we present RGRecSys, a library that allows users to perform a comprehensive robustness evaluation for recommender systems. Our library is open-source and can be accessed at \url{https://www.github.com/salesforce/RGRecSys}. 
Our future work aims to extend our library functionality to evaluate models under other types of attacks and transformation, biased and noisy interactions,
and the cold-start issue \cite{li2019zero}. In addition, we aim to incorporate fairness \cite{li2021user} and explainability \cite{zhang2020explainable} metrics to our current metrics, to allow users to test
whether a model is fair and transparent
\cite{singh2018fairness,chen2021neural}.
\bibliographystyle{ACM-Reference-Format}
\bibliography{ref}
\appendix

\end{document}